\documentclass[twocolumn, superscriptaddress,showpacs,amsmath,amssymb]{revtex4}
\usepackage{mathrsfs}
\usepackage{hyperref}

\usepackage[dvips]{graphics,epsfig}
\usepackage{amstext}
\usepackage{amssymb}
\usepackage{amsmath}
\usepackage{graphicx,color}
\usepackage{bm}

\newcommand{\s}{\mbox{\tiny S}}
\newcommand{\tS}{\mbox{\tiny S}}

\newcommand{\B}{\mbox{\tiny B}}

\newcommand{\SB}{\mbox{\tiny SB}}

\newcommand{\M}{\mbox{\tiny M}}
\newcommand{\D}{\mbox{\tiny D}}
\newcommand{\tL}{\mbox{\tiny L}}
\newcommand{\tR}{\mbox{\tiny R}}
\newcommand{\ti}{\Tilde}
\newcommand{\wti}{\widetilde}
\newcommand{\nl}{\nonumber \\}

\newcommand{\Sec}[1]{Sec.\,\ref{#1}}
\newcommand{\App}[1]{Appendix\,\ref{#1}}
\newcommand{\be}{\begin{equation}}
\newcommand{\ee}{\end{equation}}
\newcommand{\bea}{\begin{eqnarray}}
\newcommand{\eea}{\end{eqnarray}}
\newcommand{\bsube}{\begin{subequations}}
\newcommand{\esube}{\end{subequations}}
\newcommand{\ben}{\begin{equation*}}
\newcommand{\een}{\end{equation*}}
\newcommand{\Eq}[1]{Eq.\,(\ref{#1})}

\newcommand{\Fig}[1]{Fig.\,\ref{#1}}

\newcommand{\dg}{\dagger}
\newcommand{\la}{\langle}
\newcommand{\ra}{\rangle}

\newcommand{\La}{\big\la}
\newcommand{\Ra}{\big\ra}

\newcommand{\Opm}{\hat O_{\mbox{\tiny $\pm$}}}

\begin{document}

\title
{Phase-controlled quantum transport signatures in a quantum dot-Majorana hybrid
ring system}

\author{Sirui Yu}
\affiliation{School of Physics, Hangzhou Normal University,
Hangzhou, Zhejiang 311121, China}

\author{Junrong Wang}
\affiliation{School of Physics, Hangzhou Normal University,
Hangzhou, Zhejiang 311121, China}

\author{Huajin Zhao}
\affiliation{School of Physics, Hangzhou Normal University,
Hangzhou, Zhejiang 311121, China}

\author{Hong Mao} 
\affiliation{School of Physics, Hangzhou Normal University,
Hangzhou, Zhejiang 311121, China}

\author{Jinshuang Jin} \email{jsjin@hznu.edu.cn}
\affiliation{School of Physics, Hangzhou Normal University,
Hangzhou, Zhejiang 311121, China}

\date{\today}

\begin{abstract}
 We investigate the quantum transport in a hybrid ring system 
consisting of a quantum dot (QD) coupled to two Majorana bound states (MBSs) hosted
 in a topological superconducting nanowire, threaded by a magnetic flux.  
Utilizing the dissipaton equation-of-motion approach, 
we demonstrate that the differential conductance shows periodic behavior and 
 its periodicity depends on both the QD energy level and the MBS overlapping.
 A zero-bias peak (ZBP) emerges as a result of  
  the balance between normal   
 and anomalous tunneling processes, 
 associated with the presence of
a single MBS.
  Beyond the phase-dependent periodic behavior,
  the shot noise exhibits voltage-dependent transitions 
  between sub-Poissonian ($F = 0.5$), 
Poissonian ($F = 1$),  
and super-Poissonian ($F > 1$) regimes. 
Strikingly, we find a giant Fano factor ($F\gg1$) 
emerging at the balance point, accompanied by a peak in the shot noise.
This distinctive feature may serve as a supplementary signature for MBS detection. 
However, both ZBP in the differential conductance and shot noise peak
 are degraded by thermal effects.

\end{abstract}

\maketitle

\section{Introduction}
Inspired by the pioneering theoretical contributions of Kitaev in 2001 \cite{Kit01131}, Majorana bound 
states (MBSs) have garnered widespread attention in condensed matter physics
and quantum information science. 
On the one hand, MBSs possess distinctive physical attributes, 
characterized by their non-Abelian braiding statistics \cite{Kit032}.
On the other hand, owing to these exceptional properties, 
MBSs hold promise for applications in topological quantum computers
\cite{Kit01131,Kit032,Nay081083,Ali10125318,Lei11210502,Aas16031016}.
%

 Various physical platforms have been proposed to realize MBSs  
\cite{Lut10077001,Ore10177002,
Fle10180516,Sar16035143,Moo18165302,Law09237001, 
Dan20036801,Men20036802}.
Among these, a popular scheme is
  the spin--orbit coupling
semiconductor nanowire proximity coupled to a superconductor.
In this platform, %
zero-energy MBSs, also referred to as Majorana zero-energy modes (MZMs),
 are expected to reside at the nanowire ends
\cite{Lut10077001,Ore10177002}.
Current experimental efforts focus on  
  detecting and verifying MBS signatures,
  with transport measurements serving as a key diagnostic tool.
For instance, distinct transport signatures have been demonstrated in setups where
  MBSs are connected to either single lead (two-terminal) 
\cite{Fle10180516,Sar16035143,Moo18165302,Law09237001}, 
or two leads (three-terminal) \cite{Dan20036801,Men20036802}, typically via
  the observation of zero-bias peak in the tunneling conductance spectroscopy.
In addition, hybrid architectures incorporating quantum dots (QDs) as intermediaries between MBSs 
and transport leads, 
 have been proposed \cite{Cao12115311,Zoc13036802,Fen21123032,Liu15081405,Smi17063020,
 Fen22035148,Smi22205430,Smi19165427,Cao23121407,Zha25023006}. 
 These studies suggested that
 Majorana signatures may also be inferred through 
 current noise measurements.
 Although these measurements provide encouraging hints,
  the conclusive evidence for MBSs have not yet been obtained,
and the distinctive transport characteristics associated with their presence
 require further exploration.

In this context, the study of the nonequilibrium transport through a 
 hybrid ring system involving MBSs 
 represents an exciting and promising avenue of research.
The system consists of a QD coupled 
 to two Majorana modes located at the ends of a topological
 superconducting nanowire (TNW), with a magnetic flux $\Phi$ through the loop,
  as schematically shown in \Fig{fig1}. 
 This hybrid ring configuration was initially proposed by Flensberg \cite{Fle11090503} for 
 performing the non-Abelian operations on MBSs through magnetic flux tuning.
Subsequently, Liu and Baranger \cite{Liu11201308} 
predicted that this system can be used to tune Flensberg’s
qubit to the required degeneracy point (the presence of single MBS), where 
the zero-temperature peak conductance reaches half of the quantized value,
$G=e^2/h$, while vanishing elsewhere.
 This half-quantized conductance has been demonstrated to be 
 robust against variations in the QD energy level \cite{Ver14165314}
and has been measured in a QD coupled to an InAs--Al nanowire heterostructure \cite{Den161557}.
The QD--TNW interferometer can also serve as a mesoscopic probe of interwire couplings
\cite{Ram18045301}.

In recent years, the hybrid QD--TNW ring system has garnered significant interests 
 \cite{Wan181,Zen17127302, Cal201900409,Smi21205406, Smi24195410,Gon22122,Mat22155409,Mat231616,Bah25045920}. 
A number of signatures associated with MBSs have been identified in the strong coupling regime, such as
  Fano resonances \cite{Wan181,Zen17127302, Cal201900409},  
  fractional entropy \cite{Smi21205406}, and differential finite-frequency quantum noise \cite{Smi24195410}.
Furthermore, studies have addressed Coulomb-interaction-enhanced 
double-resonant tunneling \cite{Gon22122}, 
and the influence of electron–phonon coupling on conductance 
periodicity \cite{Mat22155409,Mat231616}. 
 More recently, a four-MBS configuration with two TNW loops has been investigated, 
 uncovering periodic structures in both linear and differential conductance \cite{Bah25045920}. 
 Despite these advances, the transport properties originating purely MBSs
  are still not fully understood, even in the simple geometry of \Fig{fig1} 
  where electron–electron and electron–phonon interactions are neglected.

In this work, we present a comprehensive study
of the magnetic-flux-dependent differential conductance and shot noise
 in the hybrid QD--TNW ring interferometer (\Fig{fig1}).
We focus on the weak coupling between the QD and electrodes 
at finite temperature---a scenario readily accessible in current experiments. 
Our results show that the signatures originating from MBSs are modulated by the flux.
Accordingly, the differential conductance and shot noise display
the periodic structures.
We provide a general condition governing the periodicity. 
Notably, the shot
noise reveals voltage-dependent statistical regimes:
sub-Poissonian (Fano factor $F=0.5$), Poissonian ($F=1$), and super-Poissonian ($F>1$).
A pronounced Fano factor emerges at zero bias, originating purely from MBSs.
All these transport properties are thoroughly
investigated in our study.
We believe these results contribute to deeper understanding 
of MBS-mediated quantum transport 
through mesoscopic systems.

The present numerical calculations are performed using 
 the fermionic dissipaton equations-of-motion (DEOM)
approach, which provides accurate evaluations on the 
 transport current and its noise spectrum \cite{Yan14054105,Jin15234108,Yan16110306}.
DEOM is a quasi-particle generalization of
the well--established hierarchical equation of motion
(HEOM) \cite{Tan89101,Tan906676,Tan06082001,Tan20020901,
Xu05041103,Jin08234703}. HEOM rooted in the path integral influence functional theory
\cite{Fey63118,Wei08,Kle09}, to efficiently simulate interacting open quantum systems. 
Its hierarchical construction captures dissipation, memory time,
Columbic interactions, and cotunneling.
The key features for fermionic HEOM/DEOM are as follows
\cite{Jin08234703,Zhe121129,Li12266403,Ye16608}:
(1) For noninteracting systems,
  the second-tier truncation 
is exact and agrees with Landauer–Büttiker scattering
  \cite{Dat95} and nonequilibrium Green function methods \cite{Hau08} and
(2) for interacting systems, convergence is rapid with finite tiers,
which depends on both the system and bath configurations.
HEOM/DEOM has been widely applied transient
 transport \cite{Zhe08184112,Zhe08093016,Zhe09164708,Zhe13086601},
thermopower \cite{Ye14165116}, and the spectral density of local impurity system in
the Kondo regime \cite{Li12266403,Wan13035129}.
Via the dissipaton decomposition for the hybridizing bath (electrodes),
 DEOM theory comprises
both the HEOM formalism \cite{Jin08234703}
and the dissipaton algebra that addresses
the hybrid bath dynamics.
Consequently, DEOM serves as a unified framework for evaluating various noise-related
spectra \cite{Zha18780,Wan20041102,Zha16237,Zha16204109},
including the quantum current noise \cite{Jin15234108,Jin20235144}.
For further detail, we refer to Refs.\cite{Yan14054105,Jin15234108,Yan16110306}.

The remainder of this paper is organized as follows. In \Sec{thmeth}, we give
the model description and introduce the definitions 
of the current and its power spectrum. Based on the DEOM method 
which is briefly introduced
in tbe \App{appdeom},
we carry out the precise
numerical results in \Sec{thresu}.
First,
we analyze the QD--wire Hamiltonian eigenspectrum in \Sec{thham}. 
Then we demonstrate the detailed results
for the differential conductance and shot noise \Sec{thcond} 
and \Sec{thshot}, respectively. Finally,
we give the summary in \Sec{thsum}.

\section{Methodology}
\label{thmeth}
 
\begin{figure}
\includegraphics[width=0.65\columnwidth]{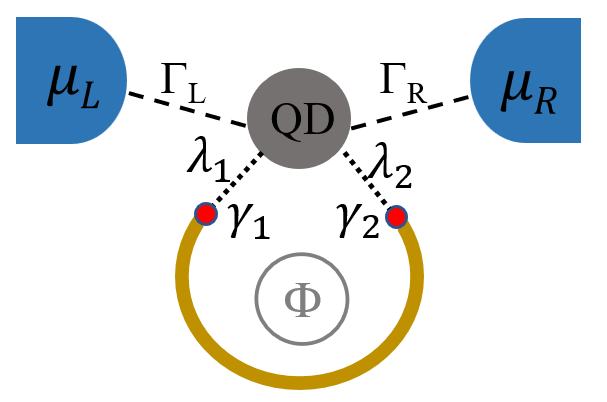} 
\caption{
Schematic diagram for the transport through the QD-wire hybrid ring system.
The QD is contacted by the two electron reservoirs under the bias voltage
($V=\mu_{\tL}-\mu_{\tR}$) with the tunneling rates $\Gamma_{\rm L}$ and $\Gamma_{\rm R}$. 
It is further side coupled to the two Majora modes ($\gamma_1$ and $\gamma_2$) 
at the two ends of a 1D topological superconductor nanowire
with the coupling coefficients $\lambda_1$ and $\lambda_2$, respectively. 
$\Phi$ is the magnetic flux through the
loop.
  }
\label{fig1}
\end{figure}
We investigate quantum transport through 
a hybrid ring system comprising a quantum dot (QD) coupled to two Majorana 
modes ($\gamma_1$ and $\gamma_2$), located at the ends of 
a topological superconducting nanowire (TNW),
as illustrated in \Fig{fig1}.
The total Hamiltonian, $H_{\rm tot}=H_{\s}+H_{\B}+H_{\s \B}$, consists of three 
components.
The central QD-TNW hybrid system is described by the 
low energy Hamiltonian \cite{Fle11090503,Liu11201308},
\begin{align}\label{HS0}
  H_{\s}=\varepsilon_{\D}\hat d^\dg \hat d \!+\! \frac{i}{2} \varepsilon_{\M}\hat\gamma_{1}\hat\gamma_{2} 
  \!+\! (\lambda_{1}\hat d^\dg\hat\gamma_{1}\!+\! i\lambda_{2}\hat d^\dg\hat\gamma_{2}+{\rm H.c.}).
 \end{align}
The first term represents the Hamiltonian of the QD,
characterized by a single fermionic mode with creation 
(annihilation) operator $\hat d^\dg$ ($\hat d$)
 and the electronic energy level $\varepsilon_{\D}$.
The second term describes the TNW with
 a pair of Majorana
modes $\hat\gamma_1$ and $\hat\gamma_2$, and their coupling energy $\varepsilon_{\M}$,
which depends on the wire length.
 The last term accounts for the coupling Hamiltonian between the QD and two Majorana modes,
  with coupling coefficients $\lambda_{1}$ and $\lambda_{2}$.
 Without loss of generality, 
 we set
  $\lambda_{1}=\lambda $ to be real and $\lambda_{2}=\lambda e^{i\phi}$, assuming
   symmetrical coupling strengths.
 The phase difference between the two couplings ($\phi=\arg(\lambda_{2}/\lambda_{1})$),
 is related to the magnetic flux $\Phi$ via $\phi=\Phi/\Phi_{0}$,
 where $\Phi_{0} = h/e$. 
%

By expressing the Majorana operators in terms of regular fermionic operators via
  $\hat\gamma_{1}=\hat f +\hat  f^\dg $ and $\hat\gamma_{2} = -i(\hat f -\hat  f^\dg)$,
 where $\hat  f^\dg$ ($\hat f $) is the fermionic creation (annihilation) operator, 
the Hamiltonian \Eq{HS0} can be rewritten as
\begin{align}\label{HS1}
H_{\s}&=\varepsilon_{\D}\hat d^\dg \hat d + \varepsilon_{\M}\hat f^\dg \hat f&\nl
&\!+ \!\lambda\big[(1\!+\!e^{i\phi})\hat d^\dg \hat f \!+\! {\rm H.c.}\big]
\!+\! \lambda\big[(1\! -\! e^{i\phi})\hat d^\dg \hat f^\dg \!+ \!{\rm H.c.}\big].
\end{align}
In this form,
the coupling between QD and MBSs consists of two distinct processes:
the normal tunneling (NT) term, given by 
$\lambda\big[(1\!+\!e^{i\phi})\hat d^\dg \hat f \!+\! {\rm H.c.}\big]$,
and anomalous tunneling (AT) term, given by 
$\lambda\big[(1\! -\! e^{i\phi})\hat d^\dg \hat f^\dg \!+ \!{\rm H.c.}\big]$.
 These two tunneling mechanisms are associated with local and crossed Andreev reflection processes,
 respectively, though a detailed analysis of these 
 is beyond the scope of this work. 
Here, we focus on the distinct transport behavior arising from the competition between NT and AT,
which can be tuned by the magnetic flux phase $\phi$. The amplitude of these
processes are modulated by factors proportional to $\propto 1\! \pm\! e^{i\phi}$.
For clarity, we classify  
  the phase-dependent
  behavior into three cases: (i) $\phi=2m\pi$,
  where only one tunneling process is present,
  either NT (ia) or AT (ib); 
  (ii) $\phi= (2m+1)\pi/2$, where NT and AT are equally weighted;
  and (iii) other phase values, where NT and AT coexist with different amplitudes.
 We will explore these transport characteristics through
 the differential conductance and shot noise.

The bath Hamiltonian for the two electron reservoirs 
 is given by $H_{\B}=\sum_{\alpha}(\varepsilon_{\alpha k}-\mu_\alpha)
\hat c^\dg_{\alpha k}\hat c_{\alpha k}$, with
the applied bias voltage $V=\mu_{\tL}-\mu_{\tR}$ and $\alpha={ L,R}$. 
The system-bath coupling Hamiltonian ($ H_{\s\B}$) describes the 
standard electrons tunneling between the QD and electrodes:
  \begin{align}\label{Hcoup}
  H_{\s\B}&=  \sum_{\alpha k}\left( t_{\alpha k} \hat d^\dg \hat c_{\alpha k}+{\rm H.c.}\right).
 \end{align}
The hybridization spectral function is assumed to be Lorentzian,
\be\label{Jw}
J_{\alpha }(\omega)
\equiv\pi\sum_k t_{\alpha  k}t^\ast_{\alpha  k}\delta(\omega-\varepsilon_{\alpha k})
=\frac{\Gamma_{\alpha}W^2}{\omega^2+W^2}.
\ee 
 Throughout this work, we adopt units of $e=\hbar=1$
for the electron charge and the Planck constant.

 The current operator for the electron
transfer from $\alpha$-reservoir to
the hybrid system is
$\hat I_{\alpha}\equiv -\dot{\hat N}_{\alpha}
=i[\hat N_{\alpha},H_{\rm tot}] =-i  \big(\hat  d^{\dg} \hat F_{\alpha }
    -\hat F^{\dg}_{\alpha } \hat d \big)$,
where $\hat N_{\alpha}=\sum_{k}\hat c^{\dg}_{\alpha k}\hat c_{\alpha k}$
is the electron number operator in
the $\alpha$-reservoir.
The mean steady--state transport current is denoted by $\bar I_{\alpha}
= {\rm Tr}_{\rm tot}\big(\hat I_{\alpha}\rho^{\rm st}_{\rm tot}\big)
\equiv \la \hat I_{\alpha}\ra$
 and its fluctuation is given by,
\be\label{corr-curr}
  \La \delta{\hat I}_\alpha(t)\delta{\hat I}_{\alpha'}(0)\Ra
=\La [{\hat I}_\alpha(t)-\bar I_{\alpha}]
  [{\hat I}_{\alpha'}(0)-\bar I_{\alpha'}]\Ra.
\ee
The lead-specified current noise spectrum is the Fourier transformation,
\be\label{Sw_alp}
  S_{\alpha\alpha'}(\omega)=\int_{-\infty}^{\infty} \!dt\,
  e^{i\omega t} \La \delta{\hat I}_\alpha(t)\delta{\hat I}_{\alpha'}(0)\Ra.
\ee
In contrast to the symmetrized version,
$S^{{\rm sym}}_{\alpha\alpha'}(\omega)
=S_{\alpha\alpha'}(\omega)
+S_{\alpha'\alpha}(-\omega)$,
the asymmetric $S_{\alpha\alpha'}(\omega)$
is directly related to the absorption and emission processes
with $\omega>0$ and $<0$,
respectively
\cite{Eng04136602,Rot09075307,Yan14115411,
Bas10166801,Bas12046802,Del18041412}.
Here, we focus on the zero-frequency counterpart, referred to as shot noise,
especially the auto-correlation noise, 
$S^{\rm sym}_{\alpha\alpha}(0)=2S_{\alpha\alpha}(0)$.
The Fano factor \cite{Bla001} is defined as $F_{\alpha}=S^{\rm sym}_{\alpha\alpha}(0)/2I=S_{\alpha\alpha}(0)/I$. 
The value of this factor can be $F_{\alpha}>1$, $F_{\alpha}=1$, and $F_{\alpha}<1$,  
describing super-Poissonian, Poissonian, and sub-Poissonian distributions, respectively.


The model under consideration can be studied exactly using several well-established methods, including
 the Landauer–Büttiker scattering
theory \cite{Dat95}, nonequilibrium Green function technique \cite{Hau08}, and 
 the dissipaton equation-of-motions (DEOM) approach \cite{Yan14054105,Jin15234108,Yan16110306}.
In this study, we employ the fermionic DEOM method 
 \cite{Yan14054105,Jin15234108,Yan16110306} to perform numerically accurate evaluations of
 both the differential conductance
and the shot noise at finite temperature. A brief overview of the DEOM
methodology is provided in the \App{appdeom}.

 \begin{figure*}
\centering
\includegraphics[width=0.680\textwidth,clip=true]{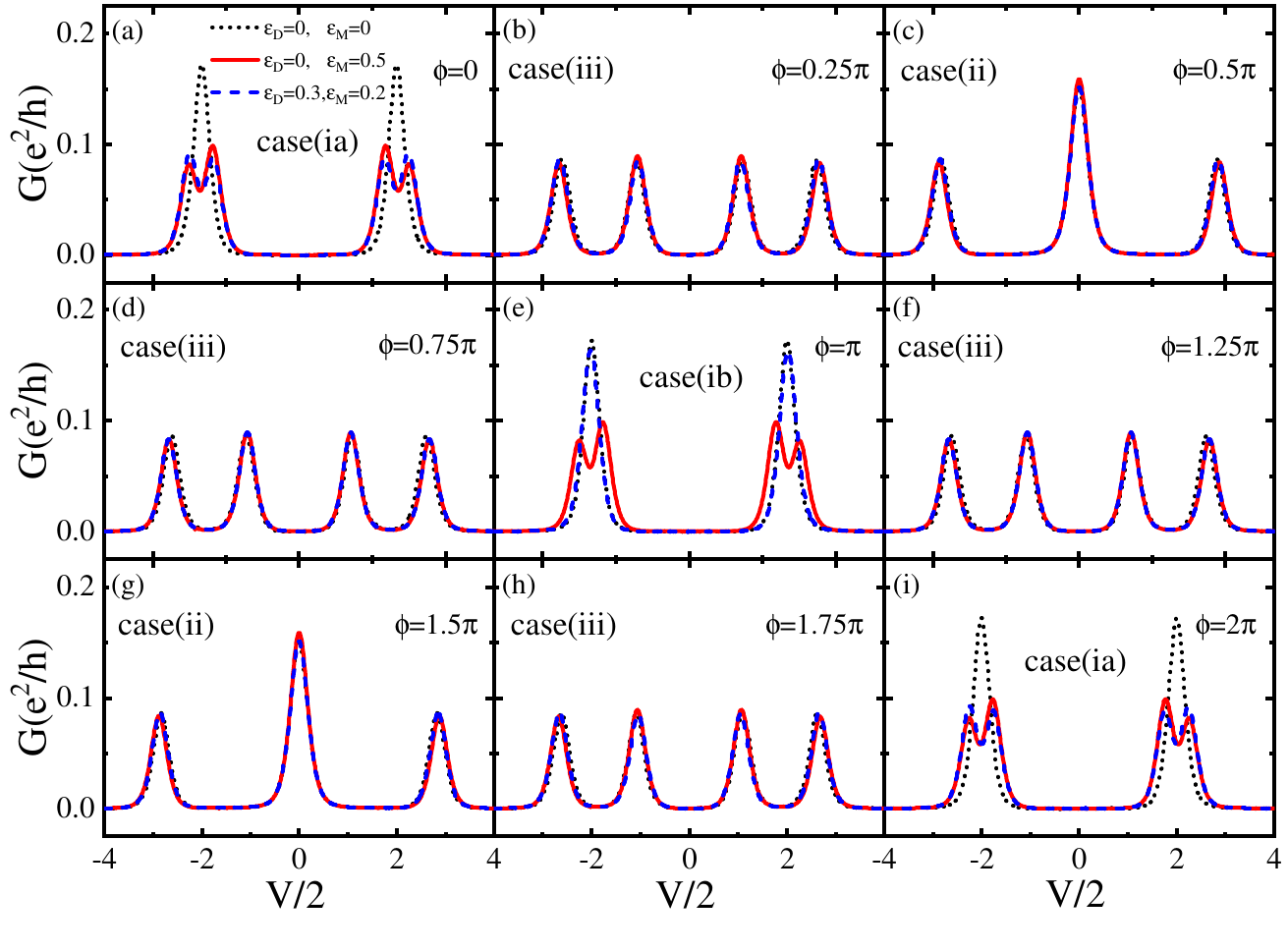}
 \caption{  Differential conductance $G$ as a function of bias voltage 
  tuned by the flux phase $\phi$
 for different $\varepsilon_{\D}$ and $\varepsilon_{\M}$. 
%
The conductance $G$ exhibits
a period of $\pi$ when $\varepsilon_{\D}\varepsilon_{\M}=0$, whereas its period is $2\pi$ 
   when
  $\varepsilon_{\D}\varepsilon_{\M}\neq0$.
   We classify the phase-dependent
  behavior into four cases: (ia) $\phi=2m\pi$, (ib) $\phi=(2m+1)\pi$, (ii) $\phi= (2m+1)\pi/2$,
  and (iii) other phase values.
  \label{fig2}}
\end{figure*} 

\section{Results}
\label{thresu}

\subsection{Hamiltonian eigenspectrum}
\label{thham}

Before presenting the results for differential conductance and shot noise,
we will first analyze the Hamiltonian eigenspectrum for the QD-wire hybrid system,
following a procedure similar to that described in our previous work \cite{Zha25023006}.
In the Fock states basis $\{|n_{\D}n_{\M}\ra=|01\ra,|10\ra,|00\ra,|11\ra\}$,
 associated with 
 $ \hat n_{\D}= \hat d^\dg  \hat d $
and $ \hat n_{\M}= \hat f^\dg \hat f $,
the Hamiltonian in \Eq{HS1} becomes block diagonal within the
subspaces of odd parity $\{ |01\ra, |10\ra \}$ (${ p}= -1$)
and even parity $\{|00\ra, |11\ra\}$ (${ p}= +1$),
\begin{align}\label{Hsm}
H_{\s}={
  \left( \begin{array}{cc}
           H_{-} & 0 \\
           0 & H_{+}
         \end{array}
         \right)},
\end{align}
where
\bsube
\begin{align}
H_{-}={
\left( \begin{array}{cc}
         \varepsilon_{\M} & \lambda(1+e^{-i\phi})\\
        \lambda(1+e^{i\phi}) & \varepsilon_{\D}
       \end{array}
       \right)},
\end{align}
\begin{align}
\label{HevenU0}
H_{+}={
\left( \begin{array}{cc}
         0 & \lambda(1-e^{-i\phi})\\
        \lambda(1-e^{i\phi}) & \varepsilon_{\D}+\varepsilon_{\M}
       \end{array}
       \right)}.
\end{align}
\esube
They indicate the intrinsic coherent Rabi oscillations in the odd and even parity subspaces
induced by the electron NT and AT processes, respectively.
The corresponding parity-dependent Rabi frequency reads
\begin{align}\label{Delta}
\Delta_{\rm p}  
&=\sqrt{(\varepsilon_{\D} + p\varepsilon_{\M})^{2}+8\lambda^{2} (1-p  \cos\phi)}.
\end{align}
It is the energy difference between the two eigenstates of
$H_{\rm p}$ ($|e_{\rm p}\ra$ and $| g_{\rm p}\ra$),
i.e., $\Delta_{\rm p}= \varepsilon_{e_{\rm p}}- \varepsilon_{g_{\rm p}}$.
The eigen-energies are $\varepsilon_{e_{\rm p}}=\varepsilon_0+\frac{ \Delta_{\rm p}}{2}$
and $\varepsilon_{g_{\rm p}}=\varepsilon_0-\frac{ \Delta_{\rm p}}{2}$,
  with $\varepsilon_0=\frac{1}{2}(\varepsilon_{\D}+\varepsilon_{\M})$.
 Apparently, the Rabi frequency is periodically modulated by the threading magnetic flux
 with $2\pi$ periodicity.
%

During the tunneling processes,
the conductance peaks
 occur whenever the Fermi surface of the electrode is on resonance with the allowed transitions 
 between the ground state of the QD-wire system and 
 an excited state with opposite parity \cite{Zha25023006,Cla17201109}. 
The corresponding peak positions are given by
\begin{align}\label{peaks}
 E_{1} &=-E_4=\frac{1}{2}(\Delta_{+}+\Delta_{-}),~~
 \nl
    E_{2}&=-E_3=\frac{1}{2}(-\Delta_{+}+\Delta_{-}).
\end{align}
This leads to the zero-bias peak (ZBP) at $\mu_{\tL} = V/2 = E_2
= E_3 = 0$, which requires $\Delta_+=\Delta_-$. This condition leads to  
energy-level degeneracy, as demonstrated in Ref.\,\cite{Liu11201308,Gon22122}.
 In terms of \Eq{Delta},
the condition can be written as
\begin{align}
\label{zbpc}
\varepsilon_{\D}\varepsilon_{\M}=4\lambda^{2}\cos\phi.
 \end{align}
Here, considering the Majorana zero-energy modes (MZMs with $\varepsilon_{\M}\rightarrow0$),
  the ZBP appears at $\phi= (2m+1)\pi/2$ [case (ii)] where the NT and AT processes
contribute equally [c.f.\Eq{HS1}].
 Consequently, the condition for the ZBP in \Eq{zbpc} reduces to
 $\varepsilon_{\D}\varepsilon_{\M}=0$.
This explains why the ZBP remains robust against the  
fluctuations in the QD energy level $\varepsilon_{\D}$ when
 $\varepsilon_{\M}=0$, as shown in Ref.\,\cite{Ver14165314}.

 In the subsequent numerical calculations,
we consider a symmetrical-lead configuration
where $\mu_{\tL}=-\mu_{\tR}=V/2$ and $\Gamma_{\tL}=\Gamma_{\tR}= \Gamma/2$.
Consequently, we denote the differential conductance as
$G=G_{\rm L}=-G_{\rm R}=\frac{{\rm d}{\bar I}_{\rm L}}{{\rm d}V}$, the steady current as
 $\bar I=\bar I_{\rm L}=-\bar I_{\rm R}$,
and the shot noise as $S= S_{\rm LL}(0)=S_{\rm RR}(0)$. 
  We focus on the scenario of weak coupling between the QD and electrodes 
    at finite temperature, i.e.,
  $\Gamma\lesssim k_{\B}T$. 
  Although the master equation approaches based on second-order perturbation 
  theory can yield at least correct qualitative results in this regime \cite{Xu22064130}, 
  we utilize the numerically accurate DEOM method to obtain precise outcomes,
  particularly for the shot noise.  
  Unless stated otherwise,
 we set $\Gamma= k_{\B}T=0.1\lambda$, use
  $\lambda=1$ as the energy unit, and
assume the wide bandwidth limit $W=50
\Gamma\gg\Gamma$ for the electron reservoirs.

\subsection{The differential conductance}
\label{thcond}

Let us first examine the characteristic of the differential conductance.
As a function of bias voltage, the differential conductance exhibits 
periodic modulation with respect to
the phase $\phi$, as numerically illustrated in \Fig{fig2}.
We observe that the periodicity of $G(V)$
 is $\pi$ when $\varepsilon_{\D}\varepsilon_{\M}=0$ and $2\pi$ 
   when
  $\varepsilon_{\D}\varepsilon_{\M}\neq0$.
The underlying physical mechanism is fundamentally 
rooted in the distinct tunneling processes 
that occur between the QD and the wire. 
These processes are precisely tuned by the flux,
as described by \Eq{HS1}, and can be further elucidated as follows.

(ia) For $\phi=2m\pi$, only electron NT processes exist
between the QD and wire. 
This leads to the intrinsic Rabi oscillation
via the electron channels associated with
 $\varepsilon_{\D}$ and $\varepsilon_{\M}$ 
in the odd parity subspace (\{$|10\ra$ and $|01\ra$\}), characterized by
the Rabi frequency $\Delta_-$ [see \Eq{Delta}].
Meanwhile, $\Delta_+=\varepsilon_{\D}+\varepsilon_{\M}$
 does not represent the Rabi frequency; rather, it denotes 
 the energy difference between the states $|11\ra$ and $|00\ra$.  
Consequently, the two conductance peaks emerge at $E_{1}=E_{2}=\Delta_-/2=2\lambda$ and $E_3=E_4=
-\Delta_-/2=-2\lambda$, symmetrically positioned around zero-bias voltage 
for $\varepsilon_{\D}=\varepsilon_{\M}=0$. Otherwise, these two peaks are split, 
  as displayed in \Fig{fig2}(a).

(ib) In contrast, for $\phi=(2m+1)\pi$,  
only AT processes occur
between the QD and wire. 
This results in the intrinsic Rabi oscillation
through the channels of 
  $\varepsilon_{\D}$ (electron) and $-\varepsilon_{\M}$ (hole)
in the even parity subspace (\{$|11\ra$ and $|00\ra$\}),
characterized by the Rabi frequency $\Delta_+$ [ see \Eq{Delta}].
Here, $\Delta_-=|\varepsilon_{\D}-\varepsilon_{\M}|$
 is not the Rabi frequency but indicates the energy difference 
 between the states of $|10\ra$ and $|01\ra$.  
Unlike case (ia), the two conductance peaks 
are positioned at $E_1=E_3=\Delta_+/2$ and $E_2=E_4=-\Delta_+/2$,
  symmetrically situated around zero-bias voltage  
for $\varepsilon_{\D}=\varepsilon_{\M}$. 
In other scenarios, these two peaks are also split, as indicated in \Fig{fig2} (e).

Notably, the models in cases (ia) and (ib) can be interpreted as two 
coupled QDs (QD-QD) \cite{Zen17127302}.
Specifically, the QD with energy $\varepsilon_{\D}$ is coupled to
another QD with energies $\varepsilon_{\M}$ and $-\varepsilon_{\M}$
for case(ia) and case(ib), respectively.
As a result, the characteristics of the differential conductance 
exhibit similarities, and they are even identical for $\varepsilon_{\M}=0$.

(ii) Of particular interest is the case where $\phi= (2m+1)\pi/2$, in which 
the coherent coupling between the QD and wire involves both
   electron NT and AT processes with equal weights.
  Rabi oscillations occur in both the odd and even parity subspaces, as indicated 
  in cases (ia) and (ib), respectively. 
The ZBP appears at $V=E_2 = E_3 = 0$ in the differential conductance,
 as displayed in \Fig{fig2} (c) and (g).
 The height of ZBP is nearly twice that of the other two peaks due to the overlap of
two peaks at zero-bias voltage. 
This characteristic is analogous to that a QD coupled with 
a single Majorana mode \cite{Zha25023006}.
Indeed, using the definition
 $\hat\gamma_{12}=\hat\gamma_1\pm\hat\gamma_2$ \cite{Fle11090503},
the Hamiltonian given in \Eq{HS1} can be rewritten as  
\begin{align}\label{HS2}
  H_{\s}=\varepsilon_{\D}\hat d^\dg \hat d  
  \!+\! \lambda(\hat d^\dg\hat\gamma_{12}+{\rm H.c.}),
 \end{align}
  where we assume $\varepsilon_{\M}=0$ for simplicity.

(iii) For other values of phase, 
both the electron NT and AT processes are also involved,
  but with different weights. 
The ZBP splits into two peaks at $E_2=-E_3$, symmetrically positioned around zero-bias voltage.
The resulting differential conductance exhibits four peaks, as illustrated in \Fig{fig2} (b), (d), (f) and (h).
This feature resembles the scenario where the QD is coupled with the quasi-MZMs 
as demonstrated in Ref.\,\onlinecite{Zha25023006}.

From the above analysis, we find that the characteristic of $G(V)$ changing with the phase
 is analogous to the variation of $G(V)$ with the topological quality factor $q$, as
 studied in Ref.\,\onlinecite{Zha25023006}. Here,
  we focus on the periodic behavior of the conductance.
Under the symmetrical bias voltage condition ($\mu_{\tL}=-\mu_{\tR}=V/2$), 
the electron channel ($\varepsilon_{\M}$) and hole channel (-$\varepsilon_{\M}$) exhibit 
symmetry around $\varepsilon_{\D}=0$.
As a result, the differential conductance features in both cases (ia) and (ib) are 
identical when $\varepsilon_{\D}\varepsilon_{\M}=0$ (either $\varepsilon_{\D}=0$ or $\varepsilon_{\M}=0$),
leading to a $\pi$-periodic $G(V)$. 
In contrast, when $\varepsilon_{\D}\varepsilon_{\M}\neq0$ (both $\varepsilon_{\D}\neq0$ 
and $\varepsilon_{\M}\neq0$) 
the symmetry between the two channels is broken, resulting in a 
$2\pi$-periodic differential conductance. 
Thus, we conclude that whether 
the period of the differential conductance, 
is $\pi$ or $2\pi$, is determined by $\varepsilon_{\D}\varepsilon_{\M}=0$
or $\varepsilon_{\D}\varepsilon_{\M}\neq0$.
 This result extends the earlier qualitative 
 picture \cite{Zen17127302,Mat22155409,Bah25045920} to a quantitative one.

We would like to note that the width of the peaks arising from the measurements is determined by
the temperature in the considered weak-coupling regime, $\Gamma\lesssim k_{\B}T$.  
Whereas the height of the peaks is affected by both the coupling strength
and temperature.
This behavior can be directly understood from the analytical expression for differential conductance,
as derived through a second-order perturbation quantum master equation (QME) approach \cite{Xu22064130}, 
consistent with the methodology presented in  
Ref.\,\cite{Zha25023006}.
Taking $\varepsilon_{\D}=\varepsilon_{\M}=0$ as an example,
     we have  
  \begin{align}\label{G00}
  G(V)&=\frac{\Gamma \beta}{2^6}
  \Big({\rm sech}^2[\beta(V/4-E_1)]+{\rm sech}^2[(V/4+E_1)]
 \nl& 
  \!+\!{\rm sech}^2[\beta(V/4-E_2)]\!+\!{\rm sech}^2[\beta(V/4+E_2)]\Big).
 \end{align}
Here $E_1$ and $E_2$ given by \Eq{peaks} are reduced to 
\bsube\label{peak0}
\begin{align}
E_1=-E_4=2\lambda\big(\cos\frac{\phi}{2}+\sin\frac{\phi}{2}\big),
\\ 
E_2=-E_3=2\lambda\big(\cos\frac{\phi}{2}-\sin\frac{\phi}{2}\big).
\end{align}
\esube
 In terms of \Eq{G00}, one can easily get the corresponding half-width of the peaks as
  $\Gamma_d=2\ln(1+\sqrt{2})k_{\B}T\approx1.763 k_{\B}T$.
  This explains why the ZBP exists at $\phi= (2m+1)\pi/2$ 
  even for $\varepsilon_{\D}\varepsilon_{\M}\neq0$,
 as long as $|\Delta_+-\Delta_-|<2\Gamma_d$. 
 Furthermore, due to the temperature-dependent width, the ZBP
 cannot serve as a reliable hallmark for the existence
of MBSs in SC wires, as demonstrated in Ref.\,\onlinecite{Zha25023006}. 
Notably, the ZBP is quite robust against the energy 
fluctuations of the electron level in the dot ($\varepsilon_{\D}$)
for ideal MZMs ($\varepsilon_{\M}=0$). 
In practical experiments situations, it is
preferable to tune the energy level of the dot to $\varepsilon_{\D}\rightarrow0$, 
as the coupling energy ($\varepsilon_{\M}$) of Majorana modes is
determined by the length of the wire.

Note that the comparison of numerical results of the differential conductance
based on QME approach given by \Eq{G00} and DEOM method 
is displayed in \Fig{appfig1} of \App{appdeom}.
The QME method captures the essential features of
of the differential conductance, at least qualitatively well when
   $\Gamma\lesssim k_{\B}T$.
The accuracy of the QME results improves as 
 the coupling strength decreases relative to the temperature.


\begin{figure*}
\centering
\includegraphics[width=0.70\textwidth,clip=true]{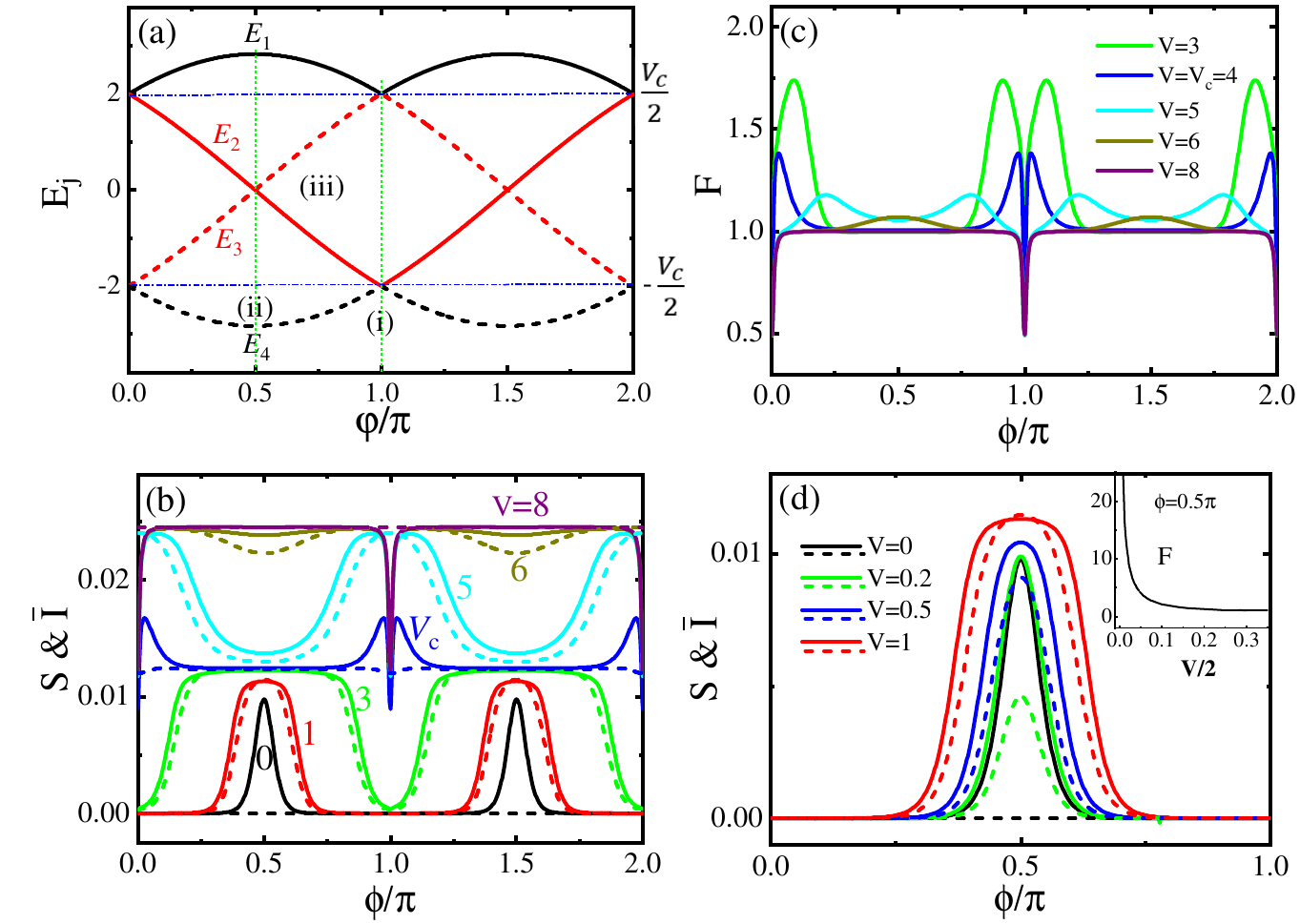}
 \caption{  Physical quantities as functions of the phase $\phi$.
 (a) Resonance energy positions $E_j$ calculated using \Eq{G00}.
 (b) Shot noise (solid line) and steady-state current (dashed line)
 under different bias voltages.
 (c) Fano factor $F=S/\bar I$.
 (d) Peak characteristics of the shot noise (solid line) and steady-state current (dashed line)
at low bias voltages.
  \label{fig3}}
\end{figure*}

\subsection{Shot noise}
\label{thshot}

 As well-known, the current noise spectrum
 contains the information much beyond the average current or conductance \cite{Her927061,Bla001,Imr02,Naz03}.
The zero--frequency noise, referred to as shot noise, primarily describes the
statistical behavior of the steady state.
It has been utilized to measure the effective carrier charge, which can be
either fraction \cite{Pic97162,Rez99238,Bid09236802} or integer
\cite{Koz003398,Lef03067002}.
The finite-frequency noise spectrum  
contains the rich correlated dynamics information
\cite{Ent07193308,Eng04136602,Jin11053704,Jin13025044,Rot09075307,Jin20235144}.  
For the system studied here, the noise spectrum would exhibit 
   coherent Rabi dip signals at $\omega\approx\pm\Delta_{\rm p}$.
  Similar to the differential conductance, the characteristics of the noise spectrum 
  are periodic with the phase, with one Rabi dip at $\Delta_-$ for case(ia) and at $\Delta_+$ for case(ib).
  In case (ii), Rabi dips degeneracy (RDD) occur at $\Delta_+=\Delta_-$,
  while, in case (iii),  
  the RDD splits into two.
  This Rabi dip feature is quite similar to that demonstrated in Ref.\,\onlinecite{Zha25023006},
  where the noise spectrum sensitive to the topological quality factor has been studied.
  Here, we do not illustrate the finite-frequency noise spectrum, but instead focus 
  on the characteristics of the zero-frequency shot noise.

 Actually, the periodic behavior of the differential conductance
 demonstrated above is also applicable 
for the current noise characteristics.
Since we focus on the underlying physical mechanism tuned by the flux, 
  we use $\varepsilon_{\D}=\varepsilon_{\M}=0$
as an illustrative example in the subsequent discussions on the shot noise.
In this example, the characteristics of the cases (ia) and (ib) are identical.
Consequently, it is unnecessary to distinguish between them,  
as both will be treated the same as case(i). 
The corresponding peak positions ($E_j$) describing the resonance transitions in the system  
are given by \Eq{peak0}.
For the convenience of discussions, we plot
 $E_j$ 
which change periodically with
the phase ($\phi$), as shown in \Fig{fig3} (a).  
Specifically, two, three, and four peaks emerge for case (i) with $\phi=m\pi$,
 case (ii) with $\phi=(2m+1)\pi/2$ , and the general case (iii), respectively.
The detailed characteristics of these conductance peaks have been demonstrated above.
Here, we denote $V_{\rm c}$ as a critical value of the bias voltage,  
at which
 the Fermi surface ($\mu_{\tL,\tR}=\pm\frac{V}{2}$) is in resonance 
 with all the energy transitions for case (i) ($\phi = n\pi$), 
 as indicated by the vertical dotted-line in \Fig{fig3} (a). 
For the parameters considered here, the critical bias voltage 
is $|V_{\rm c}|=4$.

Figure \ref{fig3} (b) displays that
the shot noise ($S$) follows the average current ($\bar I$) and 
 is periodic with the phase $\phi$, exhibiting a periodicity
  of $\pi$.
 However, their detailed characteristics depending on the phase
   is different.
This is because the generated current is only related to the opening of
 the transport channels, whereas
 the shot noise is influenced by both the transport channels and 
 charge statistical of the quasiparticles involved in the transport processes.  
We elucidate their characteristics below. 
 %

 For $|V|\geq V_{\rm c}$, which includes the two resonance transitions at $E_2$ and $E_3$, 
 the steady-state current and its shot noise exhibit a bowl-shaped profile.
 As the voltage increases, the bottom of the bowl gradually rises
 and eventually flattens. 
 This behavior is due to
the occurrence of two additional
 resonance transitions at $E_{1,4}$ for case (iii) as the bias voltage increases,
 contributing to the edge features of the bowl in \Fig{fig3} (b). 
 When the voltage reaches around $|V| = 6$, the other two resonances at $E_{1,4}$ for case (ii) also 
 become relevant. As the voltage continues to increase,
a high-voltage scenario ($|V/2|\gg |E_j|$) is reached (for instance, $V=8$) where
involves all the resonance transitions,
 the resulting steady current $\bar I$ reaches a saturated
 value regardless of the phase. 
  %

 However, the shot noise displays the dips at $\phi=n\pi$ for case (i)
 in the large bias voltage regime ($|V/2|\gg |E_j|$).
 The underlying reason is that case (i) involves
 only one tunneling processes, either NT or AT, where
 the Pauli principle comes into play during the transport process.
This lead to a sub-Poissonian distribution with the Fano factor $F=0.5$, 
as displayed in \Fig{fig3} (c). 
In contrast, in both cases (ii) and (iii), 
 the shot noise display a Poissonian distribution ($F=1$),
 indicating that the quasiparticles in the transport are bosonic  
due to the involvement of both NT and AT processes.
It is noteworthy that  
when the two resonance transitions at $E_{1}$ and $E_{4}$ occur, 
the value of shot noise is always slightly greater than the current value, 
as shown at the edges of the bowl in \Fig{fig3}(b). This results in a super-Poissonian distribution as 
illustrated in \Fig{fig3} (c), indicating the occurrence of bounching behavior at the resonances.
This behavior is enhanced with the decrease of the bias voltage.

When $|V|< V_{\rm c}=4$,  the dual resonances at $E_{1}$ and $E_4$ are fully quenched,  
resulting in an inversion of the profiles of the steady current and its shot noise,
forming an anti-bowl configuration.
For case (i) with $\phi=m\pi$ in which all transport channels are outside the 
bias window, both the steady current and the shot noise approach 
 zero at low temperatures.
In this scenario, the shot noise displays Poissonian
distribution with $F=1$ for finite temperature, which originates solely 
from the thermal fluctuations.
As a function of the phase, 
the edge features of the bowl correspond to the emergence of the
 remaining resonance transitions at $E_{2}$ and $E_3$ for case (iii). 
 Similarly, at these resonance transitions, the value of the shot noise is
 slightly greater than the current value, resulting in a Fano factor $F>1$.
 These two resonance transitions at $E_{2}$ and $E_3$ are gradually quenched 
  as the bias voltage decreases. Simultaneously, 
the current and shot noise drop to zero outside the bowl's rim,  
and the bottom of the bowl progressively narrows.

 At $V=0$ with no transport channels open, 
 the steady-state current is zero regardless of the phase. 
 Counterintuitively, the shot noise exhibits
 a peak at $\phi=(2m+1)\pi/2$ for the balance between NT and AT. This peak arises due to
  the occurrence of
 the two resonance transitions at $V=E_2=E_3=0$,
 indicating that a giant Fano factor occurs for case (ii) at zero bias voltage.
 To clearly observe this phenomenon, we specially plot \Fig{fig3} (d) 
 for very low bias voltage regime, i.e., $V\leq1$. 
 We find that as long as $V>0$, the current increases rapidly, while the shot noise grows slowly. 
When the bias reaches $V=1\gg k_BT=0.1$, indicating the onset of far-from-equilibrium situation,
  the values of the current and shot noise become nearly consistent, 
 exhibiting a Poisson distribution. 
 We further plot the Fano factor as a function of the bias voltage for case (ii)
 in the inset of \Fig{fig3} (d). It shows that the Fano factor
 decreases rapidly from a giant value to $F=1$ with the increase in the bias voltage.
 This Giant value of Fano factor at small voltages 
  may be used as an additional signature for MBSs.
However, in actual measurements, the peak at $\phi=(2m+1)\pi/2$ of 
the shot noise still has a certain broadening. 
This broadening is primarily determined by the temperature for 
the considered weak-coupling regime.  

It should be noted that  
within the QME approach, 
a compact analytic expression analogous to \Eq{G00} for 
 the differential conductance cannot be derived for the shot noise.
This stems from the significantly greater complexity involved in evaluating shot
noise compared to
 the steady-state current.
 Specifically, the steady-current is computed using 
  \Eq{Acurr}, which depends only on the steady state solution for
  $\rho^{(1)\rm st}_{j}$. In contrast, the shot noise is evaluated via
 \Eq{curr-curr}, whose solution requires ont only the steady state values from \Eq{initial_final}, 
 but also the time-dependent evolution of $ {\rho}^{(1)}_j(t;\hat I_{\alpha'})$.
 Moreover, an accurate discussion of the Fano factor ($S(0)/I$)  
 demands high numerical precision such as from DEOM theory rather than
  merely qualitative accuracy as obtained by QME approach.

\section{Summary}
\label{thsum}

In summary, we have investigated the quantum transport characteristics of a hybrid ring system 
comprising a QD coupled to MBSs in a topological
 superconducting nanowire, with a magnetic flux threading the loop.  
Using the numerically accurate dissipaton equations-of-motion (DEOM) method, 
we identified distinct magnetic-flux-dependent transport signatures. 

The differential conductance exhibits periodic behavior with respect to the magnetic flux phase, 
with $\pi$-periodicity when $\varepsilon_{\D}\varepsilon_{\M}=0$ and $2\pi$ 
  otherwise.
This modulation arises from the 
interplay between normal tunneling (NT) and anomalous tunneling (AT), 
controlled by the magnetic flux.
A prominent zero-bias peak (ZBP) appears at 
$\phi=(2m+1)\pi/2$ under the condition $\varepsilon_{\D}\varepsilon_{\M}=0$.
This is the typically MBS-contributed feature, coming from  
  the balance between the NT and AT processes. 
 This ZBP remains robust against fluctuation the QD energy level when $\varepsilon_{\M}=0$,
although its width is thermally broadened, limiting its utility as  
 a definitive signature of Majorana states.

 Similarly,
 the shot noise demonstrates phase-dependent periodic behavior.
 Beyond this feature, 
 the shot noise  
 provides insights into the statistical behavior of the transport processes.
At high bias voltages ($|V|\geq V_{\rm c}$), 
the shot noise exhibiting a bowl-shaped profile displays three voltage-dependent statistical regimes:  
(1) sub-Poissonian (Fano factor $F = 0.5$)
for the presence of only one tunneling process (either NT or AT),  
(2) Poissonian ($F = 1$), for the coexistence NT and AT processes, 
and (3) super-Poissonian ($F > 1$) at resonance edges, indicating quasiparticle bounching effects.
At low bias ($|V|<V_{\rm c}$), the quenching of resonances inverts the profiles 
into an anti-bowl shape. Notably, a giant Fano factor emerges at zero bias ($V = 0$) 
under the resonant condition ($E_2=E_3=0$) for 
the balance point. 
This anomaly rapidly diminishes with increasing voltage, 
converging to Poissonian statistics ($F = 1$) 
for $V\gg k_BT$.
The giant Fano factor at low bias may serve as a supplementary signature of MBS,
though thermal effects challenge its experimental resolution.
Our results provide a detailed and quantitative 
characterization of the interplay between MBSs, magnetic flux, 
and non-equilibrium transport in the weak-coupling regime.

\acknowledgments
 We acknowledge helpful discussions with Professor YiJing Yan and Professor Xin-Qi Li.

\appendix
\section{
DEOM theory }
\label{appdeom}

\begin{center}
 {\bf A.1 The DEOM construction}
\end{center}

 In this Appendix, we briefly introduce the DEOM method, for the details, refer to 
 Ref.\,\onlinecite{Yan14054105,Jin15234108,Yan16110306}.
We then provide the numerical comparison between exact DEOM and QME approaches
for weak system-reservoir coupling \cite{Xu22064130}.

 The system-reservoir coupling
 is generally given by
\be\label{Hsb1}
  H_{\SB}\!=\!\sum_{\alpha u }\left(t_{\alpha u k} \hat a^{\dg}_{u}  c_{\alpha k}
    + {\rm H.c.} \right)\!\!=\!\sum_{\alpha u \sigma}\hat a^{\bar\sigma}_{u}
    \wti F^{\sigma}_{\alpha u},
\ee
which accounts for multi-modes in the system (labeled by $u$) compared to the single-mode considered in \Eq{Hcoup}.
Here $ \hat F^-_{\alpha u}
=\sum_k t_{\alpha u k} c_{\alpha k}=(\hat F^+_{\alpha u})^\dg$
 and 
 $\wti F^{\sigma}_{\alpha u} \equiv \bar\sigma \hat  F^{\sigma}_{\alpha u}$,
with $\sigma =+,-$ (and $\bar\sigma$ is its opposite sign) 
for identifying the creation/annihilation operators.

 For the dissipaton description of bath interaction, \cite{Jin15234108,Yan14054105}
the bath correlation function is decomposed as
\be\label{FF_corr}
 c^{(\sigma)}_{\alpha uv }(t-\tau)\!\equiv\!\big\la \hat F^{\sigma}_{\alpha u}(t)
\hat F^{\bar\sigma}_{\alpha v}(0)\big\ra_{\B}
 \!=\! \sum_{m=1}^{M} \eta^{\sigma}_{\alpha uv m}e^{-\gamma^{\sigma}_{\alpha m}t},
\ee
using the optimal Pad\'{e} spectrum decomposition
of the Fermi function \cite{Hu10101106,Hu11244106}.
Here $\hat F^{\sigma}_{\alpha u}(t)\equiv
 e^{i H_{\B}t}\hat F^{\sigma}_{\alpha u}e^{-i H_{\B}t}$
 and $\la \cdots \ra_{\B}$ stands for the statical average over
the electrodes reservoirs in
local thermal equilibrium state.
Correspondingly, the bath operators are decomposed into quasi--particle (dissipatons),
\be\label{wtiF_f}
  \wti F^{\sigma}_{\alpha u} \equiv -\sigma \hat  F^{\sigma}_{\alpha u}
    \equiv \sum_{m=1}^{M} \hat f^{\sigma}_{\alpha u m},
\ee
which satisfy
\be\label{ff_corr}
   \big\la\hat f^{\sigma}_{\alpha u m}(t)\hat f^{\sigma'}_{\alpha' v m'}(0)\big\ra_{\B}
 =-\delta_{\sigma\bar\sigma'}\delta_{\alpha\alpha'}\delta_{mm'}\,
   \eta^{\sigma}_{\alpha u v m}\, e^{-\gamma^{\sigma}_{\alpha m} t}.
\ee
 Using composite indices
 $j\equiv(\sigma\alpha u m)$ and $\bar j\equiv(\bar\sigma\alpha u m)$,
 so that $f_j\equiv f^{\sigma}_{\alpha u m}$, etc.,
dynamical variables in DEOM are the reduced dissipaton density
operators (DDOs), which 
 are defined as,
\be\label{DDO_def}
 \rho^{(n)}_{\bf j}(t)\equiv \rho^{(n)}_{j_1\cdots j_n}(t)\equiv
 {\rm tr}_{\B}\Big[\big(\hat f_{j_n}\cdots\hat f_{j_1}\big)^{\circ}
  \rho_{\rm tot}(t)\Big]\, ,
\ee
where the product of dissipatons inside the circled parentheses, $(\,\cdots)^{\circ}$,
is \emph{irreducible}.
Exchanging any two fermionic dissipatons
introduces a minus sign, e.g.,
 $\big(\hat f_{j}\hat f_{j'}\big)^{\circ}=-\big(\hat f_{j'}\hat f_{j}\big)^{\circ}$.
 Moreover, the irreducible notation is related to the
generalized Wick's theorem, such as \cite{Yan14054105,Jin15234108,Yan16110306},
\begin{align}\label{Wick}
&\quad\, \text{tr}_{\B}\left[\big(\hat f_{j_n}\!\cdots\!\hat f_{j_1}\big)^{\circ}
    \hat f_j\rho_{\rm tot}(t)\right]   \nl
&=\rho^{(n+1)}_{j{\bf j}}
    + \sum_{r=1}^n (-)^{r-1} \La\hat f_{j_r}\hat f_j\Ra_{\B}
   \rho^{(n-1)}_{{\bf j}^{-}_r} . 
\end{align}

Followed
by using the above dissipaton algebra,
one can perform the time derivative on
  the reduced system density operator $\rho(t)\equiv{\rm tr}_{\B}[\rho_{\rm tot }(t)]$
with 
$\dot{\rho}_{\rm tot}(t)=-i[H_{\tS}+H_{\B}+H_{\SB},{\rho}_{\rm tot}(t)]$.
Consequently, the DEOM is obtained as
 \cite{Yan14054105,Jin15234108,Yan16110306}
\begin{align}\label{DEOM}
  \dot\rho^{(n)}_{\bf j}(t)&=-\bigg(i{\cal L}_{\tS}
  +\sum_{r=1}^n \gamma_{j_r}\bigg)\rho^{(n)}_{\bf j}(t)
  -i\sum_{j} {\cal A}_{\bar j}\rho^{(n+1)}_{{\bf j}j}(t)    \nl
&\quad
  -i \sum_{r=1}^n (-)^{n-r}{\cal C}_{j_r}\rho^{(n-1)}_{{\bf j}^-_r}(t),
\end{align}
which is the HEOM formalism \cite{Jin08234703}. Here,
 ${\cal L}_{\tS}\,(\cdot)=[H_{\tS},\,(\cdot)]$
and the superoperators, ${\cal A}_{\bar j}\equiv {\cal A}^{\bar\sigma}_{\alpha u m} = {\cal A}^{\bar\sigma}_{u}$
and ${\cal C}_{j}\equiv {\cal C}^{\sigma}_{\alpha u m}$
 are defined via 
\be\label{calAC}
\begin{split}
 {\cal A}^{\sigma}_{u} \Opm &\equiv
    a^{\sigma}_{ u}\Opm \pm \Opm \hat a^{\sigma}_{u}
 \equiv \big[\hat  a^{\sigma}_{u},\Opm\big]_\pm \, ,
\\
 {\cal C}^{\sigma}_{\alpha u m} \Opm  &\equiv
  \sum_{v} \big(\eta^{\sigma}_{\alpha uv m}\hat  a^{\sigma}_{v}\Opm
  \mp \eta^{\bar \sigma\,{\ast}}_{\alpha uv m}\Opm \hat a^{\sigma}_{m}\big),
\end{split}
\ee
where, $\Opm$ is an arbitrary operator,
with even ($+$) or odd ($-$) fermionic parity,
such as $\rho^{(2k)}$ or $\rho^{(2k+1)}$, respectively.
The reduced system density operator is just
$\rho(t) \equiv {\rm tr}_{\B}\rho_{\rm tot}(t)= \rho^{(0)}(t)$.
 All $\big\{\rho^{(n\geq1)}_{\bf j}\big\}$
are physically well--defined DDOs, as depicted in \Eq{DDO_def},
capturing entangled system--bath dynamics.

It is important to emphasize that 
the DEOM method offers a unified framework for studying both noninteracting 
and interacting open quantum systems, as characterized by 
the system Hamiltonian $H_{\tS}$ in the first term of \Eq{DEOM}.
For noninteracting systems, as considered here,
DEOM evaluations converge exactly
  at tier level $L=2$. 
  Indeed, we have previously derived an exact equation for the
  single-particle density matrix,  
  $\varrho_{uv}={\rm tr}_{\s}[a^\dg_v a_u\rho(t)]$,
  by truncating the hierarchy at the second tier($L=2$),
  and obtained an exact
  current expression \cite{Jin08234703}
  that agrees with the results from Landauer–Büttiker scattering theory
  \cite{Dat95} and nonequilibrium Green function method \cite{Hau08}.
For interacting systems,
DEOM exhibits rapid and uniform convergence to
the exact outcomes, as the truncated tier level $L$ (where $\rho^{(n>L)}_{\bf j}=0$) 
increases \cite{Zhe121129,Li12266403,%
Zhe13086601,Hou15104112,Ye16608}. The minimal tier $L$ required for
convergence depends on the specific configurations
of both the system and the bath.
Furthermore, the DEOM, \Eq{DEOM} forms a set of linear
differential equations, enabling the use of standard quantum dynamics pictures in
the DEOM-space descriptions of open quantum systems \cite{Yan16110306}.
This foundational structure supports consistent
DEOM-based evaluations of various physical quantities, including transport current
and its correlation function, as summarized below.

\begin{center}
 {\bf A.2 Transport current and its correlation function}
\end{center}

The transient current though the $\alpha$-lead is calculated via
$I_{\alpha}(t) = {\rm Tr}[\hat I_{\alpha}\rho_{\rm tot}(t)]$. Substituting the current operator
\be
\hat I_{\alpha}=-\dot{\hat N}_{\alpha}
=i[\hat N_{\alpha},H_{\rm tot}] 
 =-i\sum_u \big(\hat  a^{+}_u \hat F^-_{\alpha u}
   -\hat F^{+}_{\alpha u}\hat  a^-_u \big),
   \ee
    one directly obtains
\be\label{Acurr}
  I_{\alpha}(t) = -i\! \sum_{j_{\alpha}\in j}
  {\rm tr}_{\tS}\!\big[\ti a_{\bar j}\rho^{(1)}_{j}(t)\big],
\ee
where $\ti a_{\bar j}\equiv \ti a^{\bar \sigma}_{\alpha u k}
=\bar \sigma\hat a^{\bar \sigma}_{u}$
and $j_{\alpha}\equiv \{ \sigma u k\}\in j\equiv\{\sigma\alpha u k\}$.
Clearly, the transport current is directly related to the first-tier DDO.

Similarly,the lead-specific current correlation function
can be written as 
\begin{align}\label{curr-curr}
 \La\hat I_{\alpha}(t)\hat I_{\alpha'}(0)\Ra
&= 
{\rm Tr}\big[\hat I_{\alpha}\rho_{\rm tot}(t;\hat I_{\alpha'})\big]
\nl&=- i\!\sum_{j_{\alpha}\in j}
{\rm tr}_{\tS} \big[\ti a_{\bar j}
 {\rho}^{(1)}_j(t;\hat I_{\alpha'})\big],
\end{align}
which also depends on the first-tier DDOs. The key difference
lies in the initial condition: $\rho_{\rm tot}(0;\alpha')=\hat I_{\alpha'}\rho^{\rm st}_{\rm tot}$,
where $\rho^{\rm st}_{\rm tot}$ is the steady state of the whole system. 
Using dissipaton-algebra, the initial
DDOs are given by 
\begin{align}\label{initial_final}
 \rho^{(n)}_{\bm j}(0; \alpha')
&\!=\! -i\!\!\!\sum_{j'_{\alpha'}\in j'}{\ti a}_{\bar j'} \rho^{(n+1);{\rm st}}_{{\bm j}j'}
 - i\!\sum_{r=1}^n(-)^{n-r} \wti C_{j_r}\rho^{(n-1);{\rm st}}_{{\bm j}^{-}_r},
\end{align}
Further details can be found in Refs. \cite{Yan14054105,Jin15234108,Yan16110306},
along with numerical applications \cite{Jin20235144,Mao21014104}.

\begin{center}
 {\bf A.3   Weak system-reservoir coupling}
\end{center}

\begin{figure}
\includegraphics[width=1.0\columnwidth,angle=0]{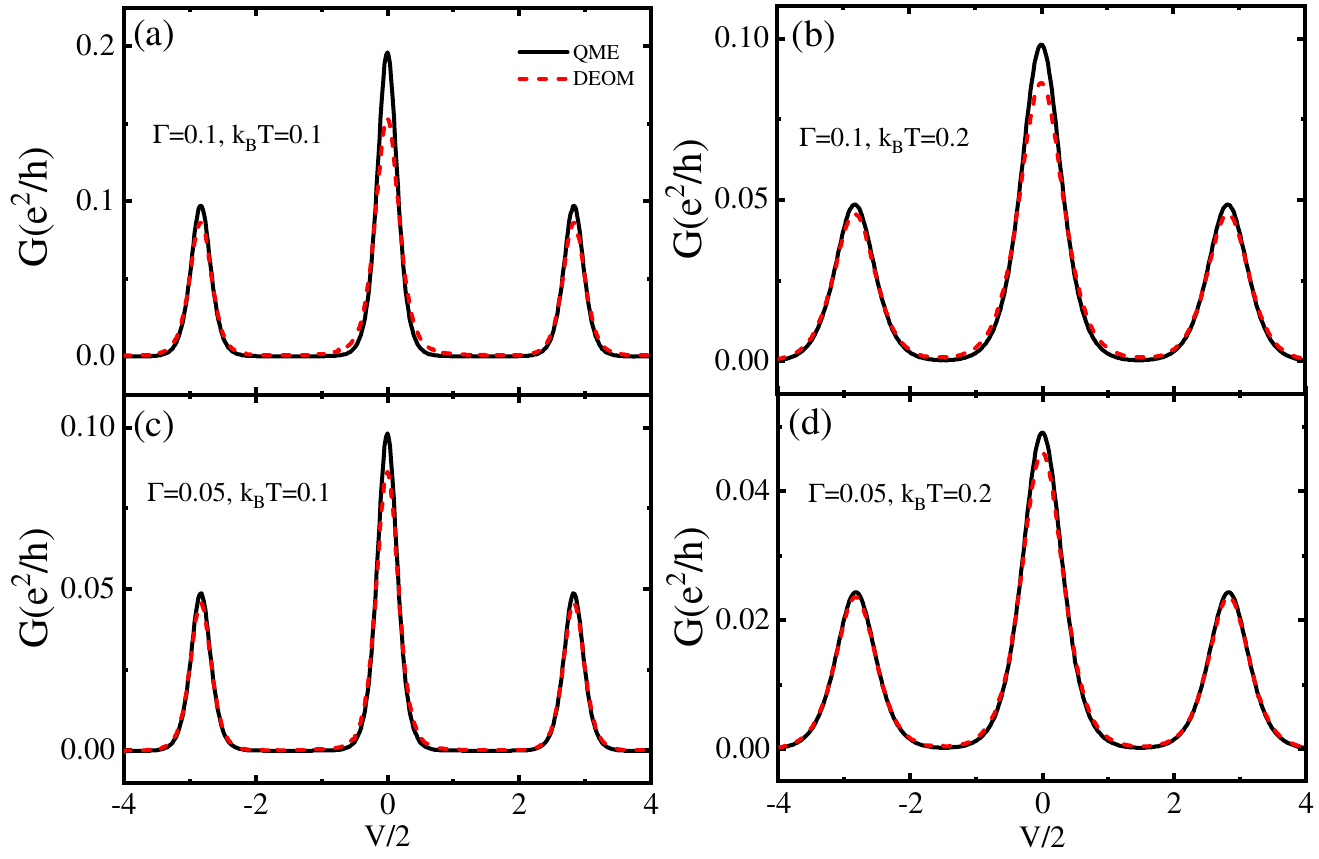}
\caption{
Differential conductance $G$ as a function of bias voltage 
for $\phi=0.5\pi$ and
  $\varepsilon_{\D}=\varepsilon_{\M}=0$. 
The evaluations are based on 
QME approach (black solid-line) given by the analytical expression
\Eq{G00} and 
exact DEOM theory (red dash-line)  
with
(a) $\Gamma= k_{\B}T=0.1$,
(b) $\Gamma=0.1$ and $k_{\B}T=0.2$,
(c)  $\Gamma=0.05$ and $k_{\B}T=0.1$, and
(d)  $\Gamma=0.05$ and $k_{\B}T=0.2$. }
  \label{appfig1}
\end{figure}

For the case of weak QD-reservoir coupling[c.f.\Eq{Hsb1}],
 the hierarchy is truncated at the lowest tier, i.e.,
 \be
 \rho^{1}_j(t)
=\rho^{\sigma}_{\alpha u m}(t)\equiv
  {\rm tr}_{\B}\big[\hat f^{\sigma}_{\alpha u m}\rho_{\rm tot}(t)\big].
  \ee
 The DEOM in \Eq{DEOM} then reduces to 
\bsube\label{TL-EOM}
\begin{align}
 &\dot\rho(t) = -i{\cal L}_{\tS}\rho(t)
 - i\sum_{\alpha u\sigma m} 
 \big[ \hat  a^{\bar\sigma}_u,\rho^{\sigma}_{\alpha u m}(t)\big],
\label{rho0t}
\\
 &\dot\rho^{\sigma}_{\alpha u m}(t)
=\!-i({\cal L}_{\tS}\!+\!\gamma^\sigma_{\alpha m})\rho^{\sigma}_{\alpha u m}(t)
  -i{\cal C}^{(\sigma)}_{\alpha u m} \rho(t).
 \label{rho1t}
 \end{align}
\esube
After straightforward algebra, we obtain 
\be\label{rho1}
 \sum_m\rho^{\sigma}_{\alpha u m}(t)
=  -i\!\int_{t_0}^t\!\!{\mathrm d}\tau\, e^{-i{\cal L}_{\tS} (t-\tau)}
 {\cal C}^{(\sigma)}_{\alpha u}(t-\tau) \rho(\tau),
\ee
where
\be\label{Ct}
 {\cal C}^{(\sigma)}_{\alpha u }(t) \bullet
 \equiv    \sum_v\big[c^{(\sigma)}_{\alpha uv}(t)\hat a^\sigma_v \bullet
 - c^{(\bar\sigma)\ast}_{\alpha uv}(t)\bullet \hat a^{\sigma}_v \big],
 \ee
with $c^{(\sigma)}_{\alpha uv }(t) $
 being the bath correlation function defined in [\Eq{FF_corr}].
 This leads directly to the well-known 
 second-order perturbation quantum master equation (QME) with memory effects
 \cite{Jin11053704,Yan05187,Mak01357,Jin16083038,Xu22064130},
\be\label{QME}
\dot\rho(t)\! =\! -i{\cal L}_{\s}\rho(t)\!-\!\sum_{\alpha \sigma u }
  \!\int_{t_0}^t\!\!{\mathrm d}\tau \big[\hat a^{\bar\sigma u},
  {\cal C}^{(\sigma)}_{\alpha u }({\cal L}_{\s},t-\tau) \rho(\tau) \big],
 \ee
where $
 {\cal C}^{(\sigma)}_{\alpha u }({\cal L}_{\s} ,t)  
 \equiv \sum_v   e^{-i{\cal L}_{\s} t} {\cal C}^{(\sigma)}_{\alpha u v}(t)$.

%

 Accordingly, the current expression given in \Eq{Acurr} simplifies to
 \cite{Jin11053704,Shi16095002,Xu22064130}
\be\label{curr-exp}
  I_{\alpha}(t)
=-\!\sum_{\sigma u} \!\!\int_{t_0}^t\!\! {\mathrm d}\tau\, {\rm tr}_{\rm s}
   [ \sigma \hat a^{\sigma u}{\cal C}^{(\bar\sigma)}_{\alpha u}({\cal L}_{\s} ,t-\tau) \rho(\tau) ].
\ee
  The stationary current is obtained by taking $t\rightarrow\infty$ and $\rho(\tau)\rightarrow\rho^{\rm st}$.
  For the current correlation function, the initial condition in \Eq{curr-curr} must be accounted for.  
The corresponding expression for the current correlation and its spectrum
can be derived similarly, and the details can be found in
Ref.\,\onlinecite{Xu22064130}.


Figure \ref{appfig1} shows the differential conductance $G$  
  computed using
the QME method (black solid-line) and the exact DEOM theory (red dashed-line).
Clearly, the QME method captures well
all the basic features of $G$ at least qualitatively for $\Gamma\lesssim k_{\B}T$.
The smaller the coupling strength is compared to the temperature, 
the more accurate the QME result becomes.

\end{document}